\documentclass[aps,preprint,floatfix,nofootinbib,showpacs]{revtex4}
\usepackage{graphicx, epstopdf}

\begin{document}

\title{Top Quark Forward-Backward Asymmetry}

\renewcommand{\thefootnote}{\fnsymbol{footnote}}

\author{ 
Kingman Cheung$^{1,2,3}$, 
Wai-Yee Keung$^4$
and Tzu-Chiang Yuan$^5$
 }
\affiliation{
$^1$Division of Quantum Phases \& Devices, School of Physics, 
Konkuk University, Seoul 143-701, Republic of Korea \\
$^2$Department of Physics, National Tsing Hua University, 
Hsinchu 300, Taiwan
\\
$^3$Physics Division, National Center for Theoretical Sciences,
Hsinchu 300, Taiwan
\\
$^4$Department of Physics, University of Illinois, Chicago IL 60607-7059, United States \\
$^5$Institute of Physics, Academia Sinica, Nankang, Taipei 11529, Taiwan
}

\renewcommand{\thefootnote}{\arabic{footnote}}
\date{\today}

\begin{abstract}
The recent forward-backward asymmetry recorded by the CDF
Collaboration for the top and anti-top quark pair production 
indicates more than $2\sigma$ deviation from the Standard Model
prediction, while its total production cross section remains
consistent.  We propose a $W'$ boson that couples to down and top
quarks.  We identify the parameter space that can give rise a large enough
forward-backward asymmetry without producing too many top and anti-top 
quark pairs.  
Other models presented previously in the literature 
that can produce such effects are also briefly discussed.
\end{abstract}


\maketitle
 
\section{Introduction} 
The recent surprise at the Tevatron came from the forward-backward 
asymmetry in top quark pair production.  
While the Standard Model (SM) only predicts a level 
as small as a few percent arising from the higher-loop contributions, 
the measurement by CDF \cite{cdf}, however, is as large as
\begin{equation}
\label{asy}
  A_{fb} \equiv \frac{N_t( \cos\theta >0) - N_t(\cos\theta < 0)}
 {N_t( \cos\theta >0) + N_t(\cos\theta < 0)} = 0.19 \pm 0.065\;({\rm stat})
   \pm 0.024\; ({\rm syst})  \;,
\end{equation}
where $\theta$ is the production angle of the top quark $t$ 
in the $t \bar t$ rest frame.  The measurement in the $p\bar p$ laboratory frame
is correspondingly smaller, because of the Lorentz boost (washout) 
of the partons along the beam axis.  
The details for event reconstruction of the $t \bar t$ pair was given in the
CDF paper \cite{cdf}.  Earlier measurement of the asymmetry with similar result 
was also reported by D$\emptyset$ \cite{D0}.

If the asymmetry is true, it will indicate the presence of new physics, 
because within the SM the asymmetry is only up to about 5\% \cite{asymm}.  
So far, a few explanations \cite{djou,mura,ferr}
have been put forward to account for such a large asymmetry.  
In Ref. \cite{mura}, an extra $Z'$ boson with a flavor-changing 
coupling $Z'-u-t$ between the top and up quarks was proposed.  
Thus, the $u \bar u$ initial state turns into the $t\bar t$ final
state via a $t$-channel $Z'$ boson exchange.  
Since the $u$ quark most of the times comes from the proton, 
the produced top quark prefers to go into the forward beam direction 
defined by the proton.

The production angle $\theta$ in the $t\bar t$ rest frame is related to
the rapidity of the $t$ and $\bar t$ in the $p\bar p$ frame by
\begin{equation}
\label{yt-costheta}
  y_t - y_{\bar t} = 2 \; {\rm arctanh} \; \left( \sqrt{ 
  1 - \frac{4 m^2_t}{\hat s} }  \; \cos\theta \right )
\end{equation}
where $\hat s$ is the square of the center-of-mass energy of the $t\bar t$ 
pair.  Therefore, the difference between the rapidities of the $t$ and
$\bar t$ in the $p\bar p$ frame is a close measure of the production angle
in the $t\bar t$ frame.  Moreover, the sign of $y_t - y_{\bar t}$ is the
same as $\cos\theta$, such that the asymmetry in Eq. (\ref{asy}) can be
given by
\begin{equation}
\label{asy-yt}
 A_{fb} \equiv \frac{N_t( y_t - y_{\bar t} >0) - N_t( y_t - y_{\bar t} < 0)}
 {N_t( y_t - y_{\bar t} >0) + N_t( y_t - y_{\bar t} < 0)}   \;.
\end{equation}
Our parton level calculation uses this definition to calculate
the asymmetry.

In this work, we propose an extra $W$-like boson $W'$ that only couples to 
the $d$ and $t$ quarks.  Thus, the $d\bar d$ initial state turns into
the $t\bar t$ final state via a charge-current exchange of the 
$W'$ boson in the $t$-channel.  
This process can be used to produce the forward-backward asymmetry
as reported by the CDF Collaboration \cite{cdf}.

Note that in the $pp$ collision, such as the Large Hadron Collider (LHC), 
the forward-backward asymmetry is lost because of the symmetric 
initial state set up.  
The Tevatron is therefore a unique place to measure the forward-backward
asymmetry for the top quark pair production.

The organization of this Letter is as follows.  In the next section, we give the 
formulas for the process that we consider.  In Sec. III, we show 
the numerical result.  In Sec. IV, we discuss our result and
other possibilities that can provide such a large asymmetry.

\section{The Formulas}

Suppose the interaction vertex for the $W'$
boson with the down and top quarks is given by
\begin{equation}
 {\cal L} = - g' \; W^{'+}_\mu \; \bar t \gamma^\mu \left( g_L P_L + g_R P_R 
\right ) d   \ + \hbox{ h.c.} \;  \;,
\end{equation}
where $P_{L,R} = (1 \mp \gamma^5)/2$ are the chirality projection operators,
$g_{L,R}$ are the chiral couplings of the $W'$ boson with fermions, and 
$g'$ is the coupling constant.

The process $ d (p_1) \; \bar d (p_2) \to t (k_1) \; \bar t(k_2)$ is 
described by two Feynman diagrams, 
one $s$-channel diagram from the one gluon exchange
and one $t$-channel diagram from the $W'$ exchange. Ignoring the $d$ quark mass,
the spin- and color-summed amplitude squared is given by
\begin{eqnarray}
\sum \left |{\cal M} \right |^2  &=& \frac{9 g'^4}{t_{W'}^2} \biggr[ 
     4 \left( (g_L^4 + g_R^4) u_t^2 
   + 2 g_L^2 g_R^2 \hat s ( \hat s - 2 m_t^2 ) \right )
+ \frac{ m_t^4}{m_{W'}^4 } 
     ( g_L^2 + g_R^2 )^2 ( t_t^2 + 4 m_{W'}^2 \hat s )      \biggr ] 
\label{5}
\\
&+& \frac{16 g_s^4}{\hat s^2} \, \left( u_t^2 + t_t^2 + 2 \hat s m_t^2 \right )
 +  \frac{16 g'^2 g_s^2}{ \hat s \, t_{W'} } ( g_L^2 + g_R^2 ) \,
         \left [ 2 u_t^2 + 2 \hat s m_t^2 
      + \frac{m_t^2}{m_{W'}^2 }( t_t^2 + \hat s m_t^2 ) \right ]
\ ,
  \nonumber 
\end{eqnarray}
where 
$\hat s = (p_1 + p_2)^2$, $t = (p_1 - k_1)^2$, $u = (p_1 - k_2)^2$ 
and
\begin{equation}
          u_t = u - m_t^2 =-\hbox{$1\over2$}{\hat s}(1+\beta\cos\theta) 
\,,\;\;\; t_t = t - m_t^2 =-\hbox{$1\over2$}{\hat s}(1-\beta\cos\theta)
\,, \;\;\; t_{W'} = t - m_{W'}^2\;,
\end{equation}
with $\beta = \sqrt{ 1 - 4 m_t^2/ \hat s}$. 
The initial spin- and color-averaged amplitude squared is given by
\begin{equation}
  \overline{\sum \left| {\cal M} \right |^2 }  
  = \frac{1}{4} \, \frac{1}{9} \; {\sum \left| {\cal M} \right |^2 }   \;.
\end{equation}
The differential cross section versus the cosine of the production angle 
$\theta$ is 
\begin{equation}
 \frac{ d \hat \sigma}{ d \cos\theta} = \frac{\beta} {32 \pi \hat s}\,
  \overline{\sum \left| {\cal M} \right |^2 }  \;,
\end{equation}
where $\hat \sigma$ denotes the cross section for the subprocess which
is then folded with the parton distribution functions to obtain the
measured cross section.
The asymmetry is obtained by integrating over the positive and 
negative range of the $\cos\theta$ variable.
In the calculation, we used $y_t - y_{\bar t}$ to calculate the asymmetry 
as given by Eqs.(\ref{yt-costheta}) and (\ref{asy-yt}).

\section{Results}

\begin{figure}[th!]
\centering
\includegraphics[width=5in]{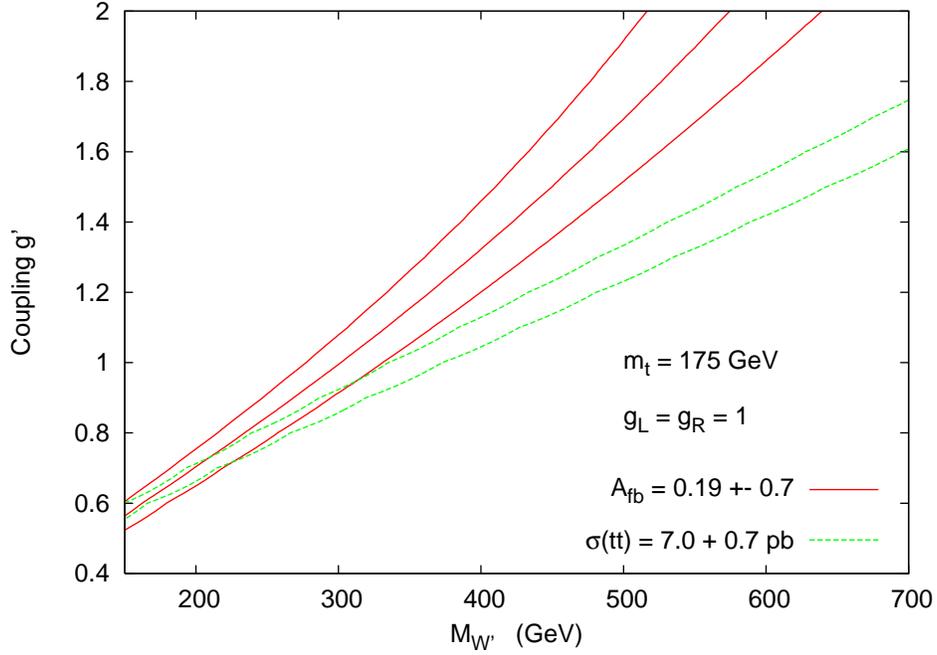}
\caption{\small The contour of the asymmetry in $t\bar t$ production
in the plane of $(M_{W'},\; g')$.  The lines shown are for 
$A_{fb} = 0.12,\; 0.19$ and $0.26$. The chiral couplings for $W' - d -t$ are 
$g_L = g_R = 1$. \label{1}}  
\end{figure}

\begin{figure}[th!]
\centering
\includegraphics[width=5in]{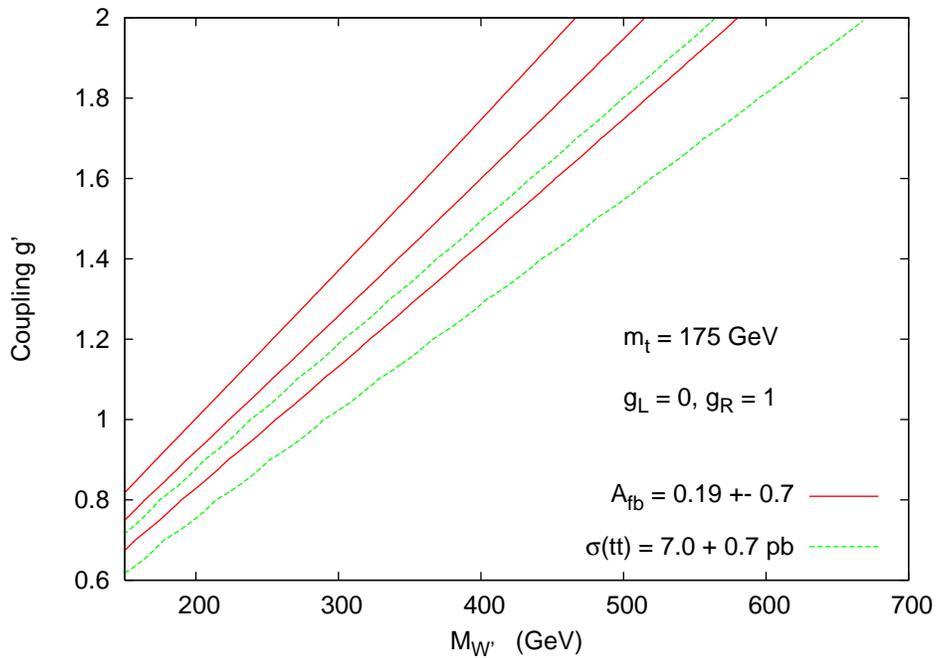}
\caption{\small The contour of the asymmetry in $t\bar t$ production
in the plane of $(M_{W'},\; g')$.  The lines shown are for 
$A_{fb} = 0.12,\; 0.19$ and $0.26$. The chiral couplings for $W' - d -t$ are 
$g_L=0,\; g_R = 1$. \label{2}}
\end{figure}

The contours of the forward-backward asymmetry are plotted in Figs. 
\ref{1} and \ref{2} for $g_L=g_R=1$ and $g_L=0, g_R=1$, respectively.
The first case is the vector-like $W'$,
while the second case is a pure right-handed $W'$.
The latter case could be a variation of certain left-right symmetric model with
a twisted family pattern that $t_R$ matches $d_R$.
The results for
these two cases are similar, because it can be seen in Eq.(\ref{5}) that
the roles of $g_L$ and $g_R$ are the same.  
The mass of the top quark $m_t$  is taken to be 175 GeV, as assumed in
the measured top quark cross section \cite{cdf-x-section}.

An important constraint is the total production cross section of the
$t\bar t$ pair.  The proposed $W'$ exchange would increase the cross
section.  We have used a tree-level 
calculation multiplied by a $K$ factor $K=1.3$, such that the tree-level
SM cross section matches the NLO result \cite{nlo}, which is about $6.8$ pb
for $m_t = 175$ GeV.  The averaged value for $t\bar t$ cross section by 
combining various channels is \cite{cdf-x-section}
\begin{equation}
 \sigma^{\rm exp} (t \bar t) = 7.0 \pm 0.3 \;({\rm stat})\; \pm 0.4 
\;({\rm syst}) \; \pm 0.4\; ({\rm lum}) \;,
\end{equation}
where the top quark mass is assumed to be 175 GeV.
We show the contour of cross sections in the plane of 
$(M_{W'}, \; g')$ in the figures.   The two contours
correspond to $\sigma(t \bar t) = 7.0$ pb (the central value) and 
$7.0 + 0.7$ pb ($+1\sigma$ value).

Combining the contours of the asymmetry under the constraint of total
cross sections, we arrive at the following conclusions:

\begin{enumerate}
\item
  In order to produce a large asymmetry either a small $M_{W'}$, 
a large coupling $g'$, or both. A heavier $W'$ boson requires a larger
$g'$ to produce enough asymmetry.  Similar effects are seen in the total 
cross section. 

\item 
  The overlapping region in each figure then gives viable parameter space
in $(M_{W'},\; g')$, which gives a large enough asymmetry without producing
too many $t\bar t$ pairs. In the first case of $g_L=g_R=1$, a $M_{W'}$
as heavy as  $300$ GeV with $g'\approx 0.8$ is allowed.  Heavier $W'$ is
not favorable. 

\item 
The second case is doing better in the sense that there are substantial 
overlapping regions between the two sets of contours.  
The mass of $W'$ can be very heavy as long as $g'$ is large enough.  
Therefore, a pure right-handed $W'$ is a favorable candidate.

\item
If $+2\sigma$ $t\bar t$ cross section is allowed, larger parameter space
can exist in the first case.

\item 
If one also takes into account the asymmetry due to higher order QCD effects,
the asymmetry required by new physics can be reduced to about $0.14$.  
Larger overlapping regions are allowed.

\end{enumerate}

\begin{figure}[th!]
\centering
\includegraphics[width=5in]{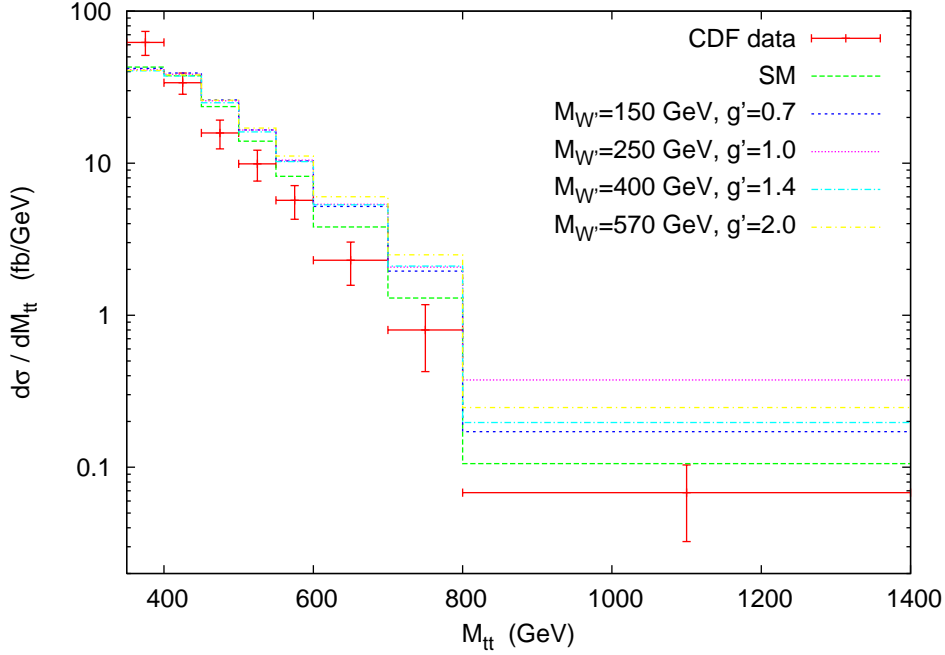}
\caption{\small 
The $M_{t \bar t}$ invariant mass distribution. Data taken from 
Table III of \cite{cdf-mtt} is shown along with the Standard Model result.
Also shown are the results of $(M_{W'} , g') = $ (150, 0.7),  (250, 1.0), 
(400, 1.4) and (570, 2). Each pair of these values are chosen within 
the $1 \sigma$ range of $\sigma (t \bar t)$ in Fig. \ref{2}.
\label{3}}
\end{figure}

Recently, the $M_{t \bar t}$ invariant mass distribution has also been published 
by the CDF \cite{cdf-mtt}.
One can use the $M_{t\bar t}$ invariant mass spectrum to put  
constraints on $g'$ and
$M_{W'}$, especially at the largest invariant mass bin as shown in  
Fig. \ref{3}.
This is because a heavier $W'$
gives more deviation for the $t \bar t$ invariant mass distribution.
However, the constraint obtained in this way correlates well with the
one obtained by the total cross section discussed erstwhile,  
though the last bin of $M_{t\bar t}$ could be stronger.
We also note that some caution must be taken if only the last bin is used.  
Experimentally, all bins in the differential cross section are correlated with one another, 
and also how the invariant masses are binned would cause strong  
bias to the results.
Fitting to the whole spectrum would make more sense.   We also noticed that the  
CDF data are more or less consistently below the SM predictions for $M_{t \bar t} > 400$  
GeV.  In fact, the same behavior was shown in the figure 4 of Ref.\cite{mura}. 

\section{Discussion}

We first discuss a few models presented in the literature 
that can account for the asymmetry.

\subsection{Non-universal $Z'$}
The flavor-changing $Z'$ scenario considered in Ref. \cite{mura} 
can be derived in a certain class of non-universal $Z'$ models
\cite{Arhrib}. The extra $Z$ boson couples to fermions in interaction basis 
with the same strength, except for the third generation.  The deviation
from family universality and thus the magnitude of FCNC are characterized
by a parameter $x$ in the $Z' - t_L -  t_L$ coupling.  The interaction of the
up sector is given by
\begin{equation}
{\mathcal L}^{(2)}_{\rm NC} =
 - g_2  Z'_\mu \; \left( \bar u, \; \bar c,\; \bar t \right)_I \; \gamma^\mu
  \left( \epsilon^u_L P_L + \epsilon^u_R P_R \right ) \; 
  \left(  \begin{array}{c}
                u \\
                c \\
                t \end{array} \right )_I
\end{equation}
where the subscript $I$ denotes the interaction basis.  The down sector is
diagonal and universal for simplicity.  For definiteness, we assume
\begin{equation}
  \epsilon_L^u = Q^u_L \left( \begin{array}{ccc}
                       1 & 0 & 0 \\
                       0 & 1 & 0 \\
                       0 & 0 & x  \end{array} \right ) \; \; \qquad {\rm and}
                   \qquad \; \; 
  \epsilon_R^u = Q^u_R \left( \begin{array}{ccc}
                       1 & 0 & 0 \\
                       0 & 1 & 0 \\
                       0 & 0 & 1  \end{array} \right ) ~.   \label{epsilonLR}
\end{equation}
When diagonalizing the up-type Yukawa coupling, we rotate
the left-handed and right-handed fields by $V_{uL}$ and $V_{uR}$, respectively.
Therefore, the Lagrangian ${\mathcal L}^{(2)}_{\rm NC}$ becomes
\begin{equation}
{\mathcal L}^{(2)}_{\rm NC} =
 - g_2  Z'_\mu \; \left( \bar u, \; \bar c,\; \bar t \right)_M \; \gamma^\mu
  \left( V^\dagger_{uL}\epsilon^u_L V_{uL} P_L + V^\dagger_{uR} \epsilon^u_R 
   V_{uR} P_R \right ) \; 
  \left(  \begin{array}{c}
                u \\
                c \\
                t \end{array} \right )_M
\end{equation}
where the subscript $M$ denotes the mass eigenbasis.  
The flavor mixing in the left-handed fields is in this case simply related to
$V_{\rm CKM}$, because $V_{dL}=1$.  Explicitly,
\begin{eqnarray}
B^{u}_L
&\equiv& V^\dagger_{uL}\epsilon^u_L V_{uL}
= V_{\rm CKM}\epsilon^u_L V^\dagger_{\rm CKM}
\nonumber \\
& \approx & Q_L^u \left(\begin{array}{ccc}
1  & (x-1) V_{ub} V_{cb}^* & (x-1) V_{ub}V_{tb}^* \\
(x-1) V_{cb} V_{ub}^* & 1  & (x-1) V_{cb} V_{tb}^* \\
(x-1) V_{tb} V_{ub}^* & (x-1) V_{tb} V_{cb}^* & x \\
\end{array}\right) \label{zput}
\end{eqnarray}
where we have used the unitarity conditions of $V_{\rm CKM}$.  By choosing 
$x\approx 0$ we have a zero $Z' - t - t$ coupling and reasonably large
$Z' - t - u$ coupling.

The forward-backward asymmetry in the top pair production arises in the 
$u\bar u \to t\bar t$ process with the $Z'$ exchange via the $t$-channel.
The formula is similarly given by Eq.(\ref{5}), with substitutions
$  m_{W'} \to m_{Z'} $,
$g'g_L \to g_2\epsilon^u_L$ and
$g'g_R \to g_2\epsilon^u_R$.

\subsection{Kaluza-Klein gluons}

Based on certain variants of the standard setup of 
Randall-Sundrum (RS) extra-dimensional model \cite{djou2}, 
the authors in \cite{djou} made use of the Kaluza-Klein (KK) states of gluon with 
different chiral couplings to the fermions to create the
forward-backward asymmetry.  The main reason for different chiral 
couplings arises from localizing the left- and right-handed
fermions in different locations in the extra dimension. 
One may also resort to the first KK gluon (without 
KK parity) or second KK gluon (with KK parity) in flat extra dimensions with different 
left- and right-handed fermion localizations \cite{csaki,park-shu,kaplan-tait}.

\subsection{Axigluon}

Without the $t$-channel exchange, the reported forward-backward
asymmetry in the top pair production disfavors flavor universal
axigluon model as shown in Ref.\cite{ferr}.  If $g_A^q = - g_A^t$
($q$ is a light quark), then an axigluon of mass from 600 GeV to 
1.4 TeV can accommodate the data.  For such a light axigluon, the LHC
experiments will certainly be able to detect it \cite{ferr}. 

\subsection{$W'$ in this work}

The $W'$ boson proposed in this work only couples to $d$ and $t$ quarks.  
It is unnatural because one normally expect the interaction to be
family diagonal or to have larger coupling with the second and third
generation fermions.  In fact, the $W' - s - t$ coupling also works but
it needs a larger coupling because of the smaller parton luminosity for strange quark.
Similarly, the $W' - b - t$ coupling should also work but with a very
large coupling.  Such large family-diagonal couplings would easily upset
existing data, e.g., $B$ meson mixings and $B$ radiative decay, unless
the $W'$ is very heavy.  The advantage of $W'- d - t$ coupling is that
there are very few existing constraints, except for the top decay into $W'$.
The top quark can decay into a $d$ quark and a $W'$ if the $W'$ is light 
enough.  Certainly, one would have seen it in the top quark 
decay at the Tevatron.  
Our analysis
showed that a heavy $W'$ works if the corresponding coupling is large 
enough, especially for the case of a purely right-handed $W'$.
In this case the $W'$ is not constrained by present top quark data.

Within QCD it was shown that \cite{asymm} higher order correction can
give an asymmetry of 5\%.  If the data   is real, then the contribution
from new physics is of order 15\%.  The QCD correction to
this new physics contribution should be very small. In other words, the excess
in asymmetry is a rather robust signature of new physics. 

If the $W'$ is not too heavy, it can be produced in the associated production
of the top quark, e.g., $gg \to W'^- t \bar d$ at the LHC.
Such a process will have a moderate cross section and the $W'^-$ will decay
into $ \bar t d$.  Unfortunately, the $t\bar t +2j$ background might be 
overwhelming.  The $W'$ can also participate in the gluon-$W'$ fusion
(similar to the $gW$ fusion for single-top production).  The resulting process
is therefore $g d \to t \bar t d$ via an intermediate $t$-channel $W'$. 
This process may increase $t\bar t$ production by a small amount at the LHC.

If the $W'$ is light enough to be pair produced and only the coupling $W' -t - q$ 
is allowed, this would also contaminate the $t \bar t$ + $2j$ background since 
both $W'^+$ and $W'^-$ would decay 100 \% into top and anti-top plus light quark jets
respectively. One possible way out of this difficulty is to include the
$W' - q -  q'$ couplings such that $W'$ would decay into light quark jets as well.
A full analysis of this scenario is interesting but beyond the scope of this work.

\section*{\bf Acknowledgments}

This research was supported in parts by the NSC
under Grant Nos. 96-2628-M-007-002-MY3 and 98-2112-M-001-014-MY3, 
by the NCTS, by the Boost Program of the NTHU,
by WCU program through the NRF funded by the MEST (R31-2008-000-10057-0),
and by U.S. DOE under Grant No. DE-FG02-84ER40173.
We would like to thank Vernon Barger for useful communication.


\end{document}